\documentclass[apj]{emulateapj}
\usepackage{apjfonts}
\slugcomment{accepted for publication in the Astrophysical Journal 
on August 5, 2007}
 
\newcommand{\lya}{\mbox{Ly$\alpha$} }
\newcommand{\msun}{\mbox{$\mathrm{M}_\odot$}}
\newcommand{\lmax}{\mbox{$L_\mathrm{Ly\alpha}^\mathrm{max}$} }
\newcommand{\lemit}{\mbox{$L_\mathrm{Ly\alpha}^\mathrm{emit}$} }
\newcommand{\lobs}{\mbox{$L_\mathrm{Ly\alpha}^\mathrm{obs}$} }
\newcommand{\fesclya}{\mbox{$f_\mathrm{esc}^\mathrm{Ly\alpha}$} }
\newcommand{\fesclyc}{\mbox{$f_\mathrm{esc}^\mathrm{LyC}$} }
\newcommand{\tlya}{\mbox{$T_\mathrm{Ly\alpha}^\mathrm{IGM}$} }
\newcommand{\xhi}{\mbox{$x_\mathrm{HI}$} }
\newcommand{\hone}{\mbox{\ion{H}{1}} }

\shortauthors{
Kobayashi,
Totani,
\& Nagashima
}

\shorttitle{
Lyman Alpha Emitters
}

\begin{document}
 
\title{
Lyman Alpha Emitters in the Hierarchically Clustering Galaxy Formation
}

\author{
Masakazu A.R. Kobayashi and 
Tomonori Totani}
\affil{Department of Astronomy, School of Science, Kyoto University,
Sakyo-ku, Kyoto 606-8502, JAPAN}
 
\and
 
\author{Masahiro Nagashima}
\affil{
Faculty of Education, Nagasaki University,
Nagasaki, 852-8521, JAPAN
}
 
 
 
 
\email{
kobayasi@kusastro.kyoto-u.ac.jp
}

\begin{abstract}
 We present a new theoretical model for the luminosity functions (LFs)
 of \lya emitting galaxies in the framework of hierarchical galaxy
 formation. We extend a semi-analytic model of galaxy formation that
 reproduces a number of observations for local and high-$z$ galaxies,
 without changing the original model parameters but introducing a
 physically-motivated modelling to describe the escape fraction of \lya
 photons from host galaxies ($\fesclya$). Though a previous study using
 a hierarchical clustering model simply assumed a constant and universal
 value of $\fesclya$, we incorporate two new effects on $\fesclya$:
 extinction by interstellar dust and galaxy-scale outflow induced as a
 star formation feedback. It is found that the new model nicely
 reproduces all the observed \lya LFs of the \lya emitters (LAEs) at
 different redshifts in $z \sim $3--6. Especially, the rather
 surprisingly small evolution of the observed LAE \lya LFs compared with the
 dark halo mass function is naturally reproduced. Our model predicts
 that galaxies with strong outflows and $\fesclya \sim 1$ are dominant
 in the observed LFs. This is also consistent with available observations,
 while the simple universal $\fesclya$ model requires $\fesclya \ll 1$
 not to overproduce the brightest LAEs. On the other hand, we found that
 our model significantly overpredicts LAEs at $z \gtrsim$ 6, and
 absorption of \lya photons by neutral hydrogen in intergalactic medium
 (IGM) is a reasonable interpretation for the discrepancy. This
 indicates that the IGM neutral fraction $\xhi$ rapidly evolves from
 $\xhi \ll 1$ at $z \lesssim 6$ to a value of order unity at $z \sim
 6$--7, which is broadly consistent with other observational
 constraints on the reionization history.
\end{abstract}
 


\section{INTRODUCTION}
 
In the cold dark matter (CDM) models of structure formation, structures
are formed and grow hierarchically via gravitational instability;
subgalactic clumps are formed in CDM halos, and they subsequently merge
and collapse to grow into more massive galaxies.  The ionizing radiation
emitted from massive stars in primeval galaxies should lead to prominent
\lya emission via the recombination of hydrogen in their interstellar
medium (ISM).  Therefore, as predicted by Partridge \& Peebles (1967),
detecting these \lya emissions via narrow- or intermediate-band imaging
is one of the most powerful technique for seeking high-$z$ young star
forming galaxies.  In the last decade, many \lya emitters (LAEs), which
seem to be powered by star formation activities, have been found through
this technique (e.g., Cowie \& Hu 1998; Hu et al. 1998; Rhoads et
al. 2000; Taniguchi et al. 2005; Shimasaku et al. 2006; Iye et al. 2006;
Murayama et al. 2007).  The \lya luminosity functions (LFs) of the LAEs
are one of the most fundamental observational quantities and they are
becoming more firmly confirmed because of the increase of the survey
fields and available samples \footnote{In this paper, the term ``LAE
LF'' refers to the \lya line luminosity function of LAEs, though some
other types of LAE LFs can be defined, e.g., UV continuum LF of LAEs.}.
Through comparison between the observed LFs and theoretical models of
the LAEs, we should be able to obtain important information for LAEs,
and more generally, for galaxy formation.
 
Nevertheless, theoretical understanding of LAEs is still developing, and
there are several different approaches for the theoretical modelling:
analytical calculations based on the halo mass function and on simple
assumptions linking \lya emission to halo mass (Haiman \& Spaans 1999;
Thommes \& Meisenheimer 2005; Dijkstra et al. 2007b; Mao et al. 2006;
Stark et al. 2007a), cosmological hydrodynamic simulations (Barton et
al. 2004; Dav\'{e} et al. 2006; Nagamine et al. 2006; Tasitsiomi 2006),
and hierarchical clustering models of galaxy formation (so-called
semi-analytic models; Le Delliou et al. 2005, 2006).  A key ingredient
of the theoretical modelling is the escape fraction of the \lya photons
from their host galaxy, $\fesclya$.  In order to predict it precisely, a
detailed calculation of radiative transfer of the \lya photons in
realistic matter distribution in a galaxy is required (e.g., Hansen \&
Oh 2006).  However, such procedure costs huge computational time because
of the resonant scatterings of \lya photons in a galaxy (Neufeld 1990;
Charlot \& Fall 1993) and a direct comparison to the observed LFs is
difficult.  Therefore, previous studies for LAE LFs introduced simple
phenomenological prescriptions.  Le Delliou et al. (2005, 2006) and
Dijkstra et al. (2007b) assumed a constant and universal escape fraction
regardless of the physical properties of galaxies.  Mao et al. (2006)
adopted screen-type interstellar dust attenuation, but a detailed merger
history of dark haloes in hierarchical clustering was not taken into
account.
 
Fortunately, there are some hints for the physical properties of LAEs
from observations.  \lya emitting galaxies in the nearby universe
generally have low metallicity or small dust amount (Charlot \& Fall
1993) and show evidence for outflow or galactic wind (Lequeux et
al. 1995; Kunth et al. 1998, 2003; Mas-Hesse et al. 2003; Keel 2005).
These properties are also found in high-$z$ Lyman-break galaxies (LBGs)
showing strong \lya emission (Pettini et al. 2002; Shapley et al. 2003;
Bower et al. 2004; Wilman et al. 2005; Frye et al. 2007; Pentericci et
al. 2007; Swinbank et al. 2007; Tapken et al. 2007).  It is
theoretically quite reasonable to expect that the amount of metal or
dust and existence of galaxy-scale outflow strongly affect the emergent
luminosity of \lya emission.  Because of the resonant scattering by
$\hone$, \lya photons have a much longer path to diffuse out from a
galaxy than UV continuum photons, and hence they are even more
vulnerable to the absorption by dust.  On the other hand, outflows would
result in a velocity difference between \hone gas and \lya photons, and
hence the escape of a \lya photon would become easier.  Therefore, these
observations and theoretical considerations indicate that the universal
and constant \fesclya in all galaxies is oversimplified, and a more
realistic modeling of \fesclya is worth investigated.
 
Another hint for the value of $\fesclya$ comes from direct observational
estimates. Gawiser et al. (2006) reported observational estimates of
\fesclya for LAEs at $z = 3.1$ as $\fesclya \sim 0.8$ (the best fit) and
an lower limit of $\fesclya \gtrsim 0.2$. However, the semi-analytic
prediction by Le Delliou et al. (2005, 2006) requires a much smaller
value of $\fesclya = 0.02$ to fit the observed LAE LFs. Therefore a new
model of LAE LFs that can reproduce the observations with $\fesclya \sim
1$ is highly desirable.
 
Here, we present a new model of LAE \lya LFs, based on a semi-analytic
model of hierarchical galaxy formation [a slightly updated version of
Nagashima \& Yoshii (2004) used in Nagashima et al. (2005)].  This
Mitaka model is one of the latest semi-analytic models in which the
merger history of dark matter haloes is taken into account and baryon
physics such as star formation and feedback is phenomenologically
treated (e.g., Kauffmann et al. 1993; Cole et al. 1994; Nagashima et
al. 1999; Somerville \& Primack 1999; Nagashima \& Yoshii 2004; Baugh et
al. 2005; Nagashima et al. 2005).  This model can reproduce most of 
the observations for photometric, kinematic, structural, and chemical
properties of local galaxies.  Moreover, the LFs and angular two-point
correlation functions of LBGs at $z=4$ and $5$ predicted by this
model coupled with N-body simulations are found to be in good agreement
with the observation (Kashikawa et al. 2006a).  We extend this model to
treat LAEs, but without changing any original parameters determined by
the fit to observations.  Rather, we try to explain the LAE LFs by
minimal extension with a least number of new and physically-motivated
parameters.  Specifically, we incorporate two new effects in the
calculation of $\fesclya$: \lya photon extinction by dust and galactic
wind driven by the supernova feedback.  The new model will be carefully
compared with all currently available data of high redshift LAE LFs.
 
The observed number of LAEs would be significantly reduced if the
intergalactic medium (IGM) is neutral, because the red wing of the
Gunn-Peterson trough should attenuate the \lya emission (e.g.,
Miralda-Escud\'{e} 1998; Haiman \& Spaans 1999).  
Therefore, the apparent evolution of LAE LF is recognized as an
invaluable probe of the cosmic reionization, and this approach has
already been applied to the observed data (Rhoads \& Malhotra 2001;
Haiman 2002; Malhotra \& Rhoads 2004; Stern et al. 2005; Kashikawa et
al. 2006; Iye et al. 2006) to constrain the reionization history, which
is complementary to the spectra of quasars (Fan et al. 2006) or
gamma-ray burst (GRB) afterglows (Totani et al. 2006).
However, a weak point of the LAE-LF method is a degeneracy
between the intrinsic evolution of the LAE LFs and the apparent
evolution by the IGM effect.  Therefore, the intrinsic LF evolution must
be known reliably to derive a robust constraint on the reionization.
However, an ad hoc assumption of no intrinsic evolution was
invoked in most of previous studies. We will discuss some implications
for reionization based on our results of LAE LF evolution.
 
The rest of this paper is organized as follows. 
In \S~\ref{section:model}, we describe our extension of the Mitaka model
to incorporate LAEs, and we compare the model results with the
observed LAE LFs at various redshifts in $3\lesssim z\lesssim 7$ in
\S~\ref{section:results}.  
We discuss implications for reionization in
\S~\ref{section:reionization}, and then summarize this work in
\S~\ref{section:conclusions}.
The background cosmology adopted in this paper is the standard
$\Lambda$CDM model:
$\Omega_\mathrm{M}=0.3,~\Omega_{\Lambda}=0.7,~\Omega_\mathrm{b}=0.04,~h=0.7,$
and $\sigma_8=0.9$.

\section{MODEL DESCRIPTION}
\label{section:model}
The detailed description of the Mitaka semi-analytic model for
hierarchical galaxy formation is given in Nagashima \& Yoshii (2004)
(see also Nagashima et al. 2005).  
Here we focus on the extension of the original Mitaka model to include
LAEs.
 
 \subsection{Star Formation Rate in Starburst Galaxies}
 
  In the original Mitaka model, all of the cold gas in a galaxy turns
  into stars and hot gas instantaneously when the galaxy is classified
  as a ``starburst'' as a result of a major merger, and hence the SFR
  cannot be defined.  
  This means that we cannot calculate the \lya photon production rate
  appropriately that is essential to determine the \lya luminosity.
  However, starburst galaxies could have a significant contribution to
  LAE LFs, because they have the highest SFR. 
  Therefore we modify the Mitaka model to calculate SFR in
  starburst populations; we adopt an exponential SFR evolution with a
  time scale of $\tau_{\rm burst}$:
  \begin{equation}
   \psi(t)=\frac{M_\mathrm{cold}^0}{\tau_\mathrm{burst}}
    \exp{\left[-\frac{t}{\tau_\mathrm{burst}}\right]}
    \label{eq-SFR} \ ,
  \end{equation}
  where $M_{\rm cold}^0$ is the initial mass of available cold gas
  for star formation. 
  We make a reasonable assumption that $\tau_{\rm burst}$ is
  proportional to the dynamical time of the newly formed spheroid, as
  $\tau_{\rm burst} = f_{\rm burst} \tau_\mathrm{dyn}$. We adopt $f_{\rm
  burst} = 10$, which is consistent with SPH simulations (e.g.,
  Kobayashi 2005) and with a recent observation of submillimeter
  galaxies at $z\sim 2-3.4$ (Tacconi et al. 2006).
 
  It is expected that not all of the cold gas will be locked up into
  stars, but some fraction of gas will be heated by supernova
  feedback, and then ejected from a galaxy as a hot galactic wind.
  This effect has already been taken into account in the Mitaka
  model, to reproduce various scaling laws of local galaxies. 
  We define the efficiency of star formation $f_*$ as $M_* = f_* M_{\rm
  cold}^0$, where $M_*$ is the finally produced stellar mass after the
  starburst episode. 
  The parameter $f_*$ is determined by the Mitaka model, and it depends
  on the dark halo circular velocity $V_c$, since the feedback should be
  more efficient for less massive (smaller $V_c$) galaxies from which
  gas can be more easily removed.\footnote{  
  The quantity $f_*(V_c)$ scales as 
  $\propto V_c^4$ and $\propto V_c^0$ 
  for $V_c \lesssim V_{c, *}$  and $\gtrsim V_{c,*}$, respectively,
  where $V_{c, *} \sim$ 130 km/s
  (Nagashima \& Yoshii 2004).}
 
  Since the galactic wind will blow up the ISM gas and stop the star
  formation, it is reasonable to identify the time of onset of the
  galactic wind, $t_{\rm wind}$, as that when the stellar mass produced
  by the above SFR evolution becomes $f_* M_{\rm cold}^0$, i.e.,
  \begin{equation}
   \int^{t_\mathrm{wind}}_0 \psi (t)\mathrm{d}t
    = f_* M_\mathrm{cold}^0,
    \label{eq-twind}
  \end{equation}
  and hence $t_{\rm wind} = - \tau_{\rm burst} \ln (1-f_*)$.
  This indicates that the galactic wind phase will onset later for
  massive galaxies, and this trend is similar to the traditional picture
  of the galactic wind in starbursts (Arimoto \& Yoshii 1987). 
  After the onset of the galactic wind, star formation is stopped and
  the interstellar gas will escape from a galaxy with a time scale of
  $\sim t_{\rm esc} = r_e / V_c$, where $r_e$ is the effective radius of
  the galaxy calculated in the Mitaka model.

  \subsection{Observed Line Luminosity of $\lya$}
  \label{subsection:obslyalum}
  Let \lemit be the \lya line luminosity emitted from a galaxy,
  which is given by
  \begin{equation}
   \lemit = \lmax
    \left(1-\fesclyc \right)
    \fesclya
    \label{eq-Lmax}
  \end{equation}
  where \fesclyc and \fesclya are the escape fraction of \hone ionizing
  photons and \lya photons from their host galaxy, respectively.  We
  adopt $\fesclyc =0$, i.e., all ionizing photons emitted from massive
  stars are absorbed in their host galaxy.  Although this assumption may
  be rather extreme, it is a reasonable treatment for our purpose
  because the observationally inferred value of \fesclyc is much less
  than the unity (Inoue et al. 2006). The modelling of \fesclya is the
  key issue of this work, and will be treated in detail in the next
  subsection.
 
  The maximally possible luminosity \lmax is the luminosity that is
  achieved if all \hone ionizing photons emitted from massive stars are
  absorbed by \hone in the ISM (i.e. $\fesclyc =0$) and then reprocessed
  into \lya photons in the ionization equilibrium.
  We define $\eta_{\rm Ly\alpha}$ as the expected number of \lya photons
  produced by one ionizing photon (or equivalently, by one
  recombination); in the case A recombination it is given by
  $\eta_{\rm Ly\alpha} = \alpha_{\rm Ly\alpha}^{\rm eff} / \alpha_A =
  0.40$ and in the case B, it becomes $\eta_{\rm Ly\alpha} = (\alpha_B -
  \alpha_{\rm 2^2S}^{\rm eff})/\alpha_B = 0.68$ (Osterbrock 1989,
  \S 4.2 and 11.8).
  Here, $T = 10^4$ K is assumed in the both cases. 
  We apply the case B following the previous studies (e.g., Charlot \&
  Fall 1993; Valls-Gabaud 1993; Haiman \& Spaans 1999; Le Delliou et
  al. 2005, 2006; Mao et al. 2006; Tasitsiomi 2006; Dijkstra et
  al. 2007b; Stark et al. 2007a), but the luminosity would not be
  changed significantly even if we apply the case A.
  Note that the case A may be appropriate for the situation of
  $\fesclya \sim 1$.
 
  Then, for a quiescent galaxy in which star formation time scale is
  much longer than the massive star lifetime,
  \lmax is given by
  \begin{equation}
   \lmax(t) = \eta_{\rm Ly \alpha} \epsilon_\mathrm{Lya}
    \psi(t) Q_\mathrm{H}^\mathrm{quies}(Z_\mathrm{cold}),
    \label{eq-SFRtoLLya-quies}
  \end{equation}
  where $\epsilon_\mathrm{Ly\alpha} = h \nu_{\rm Ly \alpha} =
  10.2$ eV and $Q_\mathrm{H}^\mathrm{quies}$ is the ionizing photon
  emission rate normalized by a unit SFR.
  On the other hand, star formation time scale could be comparable
  with the massive star lifetime in starburst galaxies, and we must take
  into account the evolution of ionizing photon production rate in
  stellar evolution. Then we calculate as
  \begin{equation}
   \lmax(t)
    = \eta_{\rm Ly\alpha} \epsilon_\mathrm{Lya}
    \int^{t}_0\psi(t')Q_\mathrm{H}^\mathrm{burst}(t-t',Z_\mathrm{cold})
    \mathrm{d}t',
    \label{eq-SFRtoLLya-burst}
  \end{equation}
  where $Q_\mathrm{H}^\mathrm{burst} (t, Z_{\rm cold})$ is the ionizing
  photon production rate from an unit stellar mass whose age is $t$.
  Note that the physical dimensions of $Q_\mathrm{H}^\mathrm{quies}$ and
  $Q_\mathrm{H}^\mathrm{burst}$ are different.
  These quantities are calculated from the result of Schaerer (2003)
  with a correction for the initial mass function used in the Mitaka
  model. 
  Both $Q_\mathrm{H}^\mathrm{quies}$ and $Q_\mathrm{H}^\mathrm{burst}$
  increase by a factor of $\sim$ 3 with decreasing stellar 
  metallicity from the solar abundance
  to the zero metallicity. The stellar metallicity is calculated from
  that of the cold gas, $Z_{\rm cold}$, in the Mitaka model.
  The evolution of $Q_\mathrm{H}^\mathrm{burst}(t)$ is characterized by
  the two phases: almost constant phase at $Q_\mathrm{H}^\mathrm{burst}
  \simeq 10^{47}~\mathrm{photons~s^{-1}~\msun^{-1}}$ for $t\lesssim
  10^6~\mathrm{yr}$ and the subsequent exponential decay with a typical
  timescale of several million years.
 
  Finally, when the IGM neutrality is high, \lya luminosity 
  could be reduced by IGM absorption, and we include this effect 
  as:
  \begin{equation}
   \lobs
    =\lemit
    \tlya ,
    \label{eq-LobsLmax}
  \end{equation}
  where \tlya is the IGM transmission to \lya photons, and it is set to
  be unity except for $z \gtrsim 6$, since $\tlya \sim 1$ has been
  established at $z \lesssim 6$ by observations (Fan et al. 2006).

  \subsection{\lya Escape Fraction Modelling}
 
  We first examine the following models for $\fesclya$:
  \begin{itemize}
   \item the simply proportional model: $\fesclya =f_0$,
   \item the dust (screen) model: $\fesclya
	 =f_0\mathrm{e}^{-x}$,
   \item the dust (slab) model: $\fesclya
	 =f_0\left(1-\mathrm{e}^{-x} \right)/x$, 
  \end{itemize}
  where $x\equiv N_\mathrm{cold}Z_\mathrm{cold}/\left(N_\mathrm{cold}
  Z_\mathrm{cold}\right)_0$, and $N_\mathrm{cold}$ is column density of
  cold gas that is computed from the cold gas mass $M_\mathrm{cold}$ and
  the effective radius $r_\mathrm{e}$ via $N_\mathrm{cold}\equiv
  \left(M_\mathrm{cold}/2\right)/\pi r_\mathrm{e}^2$.  
  The simply proportional model assumes an universal and constant value
  of \fesclya for all galaxies, as assumed in Le Delliou et
  al. (2005, 2006).
  The parameter $(N_\mathrm{cold}Z_\mathrm{cold})_0$ determines the
  strength of the absorption of \lya by dust. 
  The absorption of UV continuum by dust has already been included in
  the Mitaka model, but we treat this parameter independently for the
  \lya absorption by dust, since the effective optical depth for the
  \lya photons could be much larger than that for continuum photons
  because of much longer path caused by multiple resonant scattering of
  \lya photons by neutral hydrogen.
 
  Furthermore, we examine another \fesclya model that incorporates the
  effect of galaxy-scale outflow in addition to that of the interstellar
  dust extinction, which we call the outflow$+$dust model.  In this
  model, starburst galaxies are classified into three phases of
  pre-outflow, outflow, and post-outflow phases.  
  These phases are defined by the elapsed time $t$ from the onset of
  star formation, as $t<t_\mathrm{wind}$, $t_\mathrm{wind}\le
  t<t_\mathrm{wind}+t_\mathrm{esc}$, and $t\ge
  t_\mathrm{wind}+t_\mathrm{esc}$, respectively (see
  \S~2.1 for the definition of $t_\mathrm{wind}$ and $t_\mathrm{esc}$).
  All the quiescently star-forming galaxies are classified into the
  pre-outflow phase.  
  We model \fesclya in the outflow$+$dust model as follows:
  \begin{itemize}
   \item the pre-outflow phase: $\fesclya = f_0\mathrm{e}^{-x}$,
   \item the outflow phase: $\fesclya = f_0^\mathrm{wind}$,
   \item the post-outflow phase: $\lemit = 0$.
  \end{itemize}
  In the pre-outflow phase, we adopt the same modeling as the above dust
  models. We only apply the screen dust prescription because the
  difference between the screen and slab dust models is small, as we
  will find in \S~\ref{section:results}.  In the outflow phase,
  according to the theoretical expectation that outflow could
  drastically decrease the effective opacity for \lya photons, we adopt
  $f_0^\mathrm{wind}=0.8$ as inferred for observed LAEs (Gawiser et
  al. 2006).  Note that we do {\it not} treat
  $f_0^\mathrm{wind}$ as a free parameter, keeping the number of free
  parameters same as the dust models.
  In the post-outflow phase, while
  the \hone ionizing photons are produced at moderate rates, neutral gas
  to absorb these photons in host galaxies is absent (i.e., $\fesclyc
  =1$), and hence no \lya photons are produced ($\lemit = 0$).
 
  Therefore, the new free parameters in all the above models are $f_0$
  and $\left(N_\mathrm{cold}Z_\mathrm{cold}\right)_0$, while all other
  parameters are from the original Mitaka model without changing their
  values.

  \subsection{Luminosity Function of LAEs}
  Now we can calculate the \lya LF of LAEs by calculating $\lobs$ as
  explained above from the Mitaka model of galaxy formation.
  We determine the free parameters of \fesclya models by fitting the
  model \lya LF with $\tlya = 1$ to the data at $z=5.7$ reported by Shimasaku
  et al. (2006), which is corrected for both detection completeness and
  contamination, being the most reliable LAE LF to date.  
  Specifically, we perform a $\chi^2$ test in the whole observational
  range of the \lya luminosity ($\log{\left[L_\mathrm{Ly\alpha} /
  (h^{-2}~\mathrm{ergs~s^{-1}})\right]} = 42.1-43.1$).  
  Because the LF error bars are the smallest at the faint end, the free
  parameters of \fesclya are determined mainly by matching to the
  faint-end of the \lya LF of Shimasaku et al.
  We fix these values of the free parameters at $z = 5.7$ and then apply
  the model to the data at other redshifts with keeping the parameter
  values, because they should reflect the physical properties of
  \lya photons that are not expected to evolve with redshift.
 
\begin{figure}
 \epsscale{1.0}
 \plotone{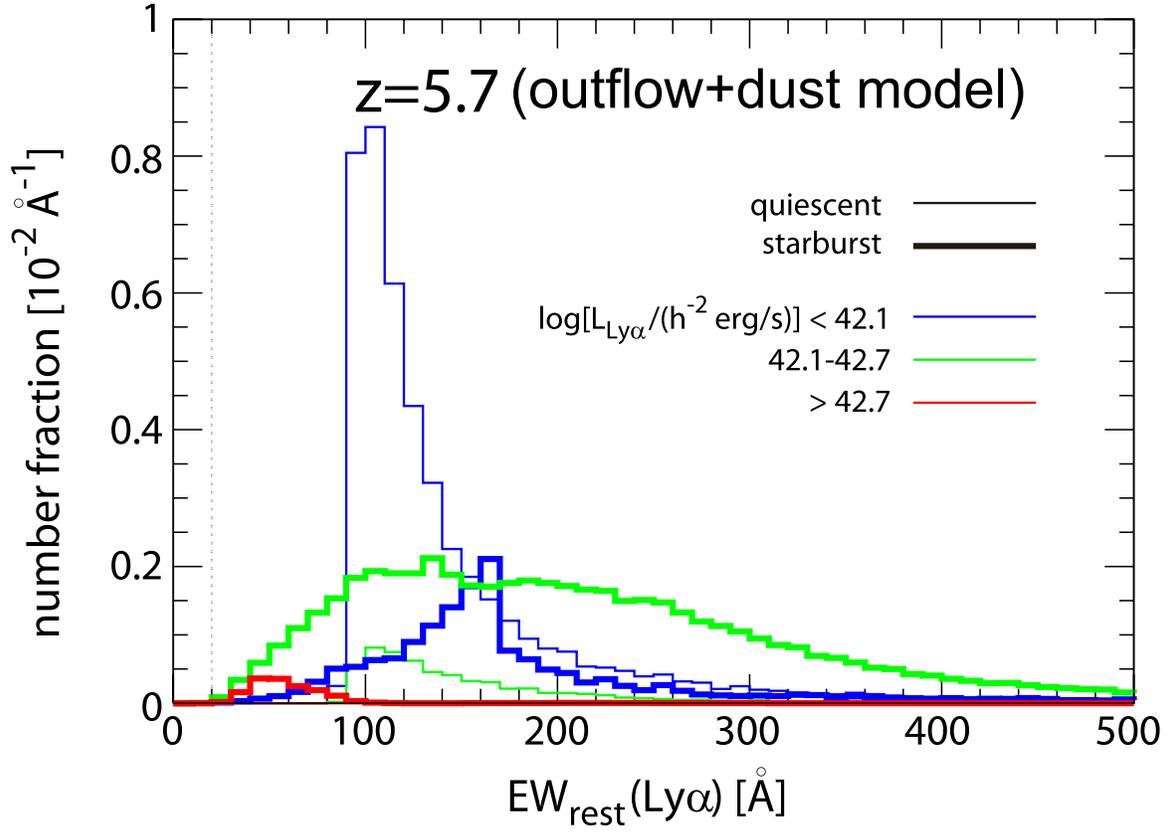}
 \caption{
 The predicted rest-frame EW of the \lya emission line for the LAEs at
 $z=5.7$ in the outflow$+$dust model, decomposed into
 three intervals of Ly$\alpha$ luminosity. The EWs are intrinsic
 without IGM absorption for \lya or UV continuum flux.
 The thin and thick histograms are the EW distributions for quiescent
 and starburst galaxies, respectively.
 The vertical dotted line represents a typical threshold 
 for galaxies to be selected as LAEs in observation.
 }
 \label{fig-EWz5p7}
\end{figure}

  It should be noted here that we do not set any criterion on the
  equivalent width (EW) of \lya lines for model galaxies to be selected
  as LAEs.  
  Though the observations of LAEs usually set threshold values for EW in 
  the selection, the effect of different EW thresholds on the predicted
  \lya LF is small.  
  We show the model calculation of \lya EW distribution of LAEs at $z =
  5.7$ in Figure \ref{fig-EWz5p7}, which is intrinsic EW and 
  the IGM absorption effect on the continuum is not taken into account. 
  Here, we calculated the EW by \lobs and the continuum luminosity 
  of model galaxies.
  \placefigure{fig-EWz5p7}
  Almost all LAEs have larger EWs than a typical threshold in
  observation, i.e., $\mathrm{EW^{rest}(Ly\alpha)}=20$~\AA, and hence our
  results do not change even if we set a threshold \lya EW used in
  observations.  
  This result is consistent with that of Le Delliou et al. (2006).
 
 \section{RESULTS} 
 \label{section:results}
 
  \subsection{Luminosity Functions at $z \lesssim$ 6}
\begin{figure}
 \epsscale{0.5} 
 \plotone{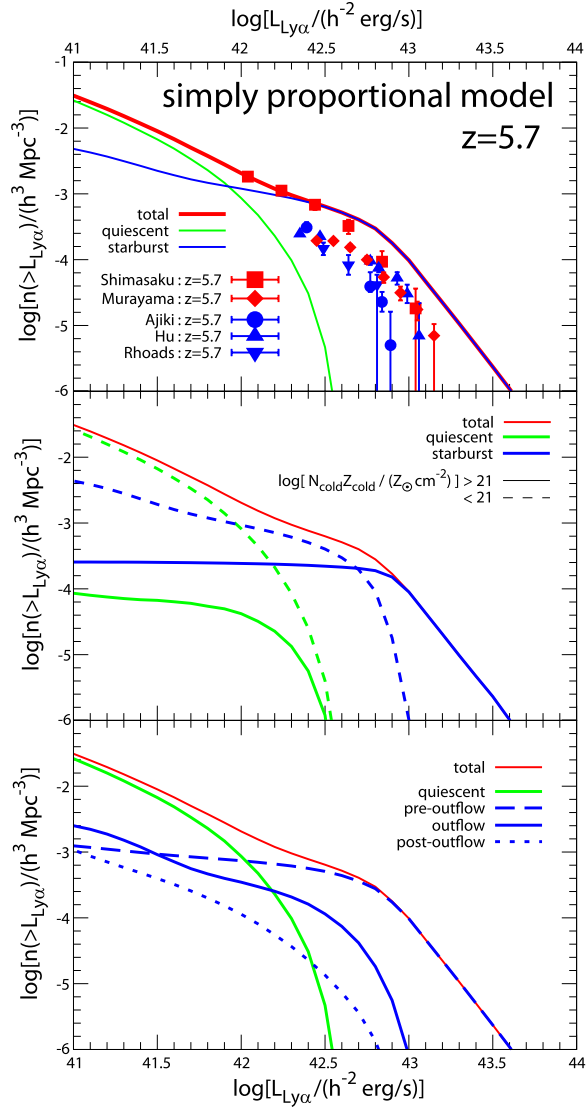} 
 \caption{ 
 The cumulative \lya LF of the simply proportional model at $z=5.7$
 with $f_0=0.60$.  
 \textit{Top}: The contributions from quiescent and starburst galaxies
 are shown separately, as indicated in the figure.  
 The symbols with error bars are observational data of the LAEs at
 $z=5.7$. The references for the data points are as follows: Shimasaku
 et al. (2006), Murayama et al. (2007), Ajiki et al. (2003), Hu et
 al. (2004), and Rhoads et al. (2003).
 Note that these observational data are not corrected for detection
 completeness and contamination, except for Shimasaku et al. (2006).
 \textit{Middle}: starburst (thick lines) and
 quiescent (thin lines) galaxies are further 
 decomposed into two groups corresponding
 to high (solid lines) and low (dashed lines) values of the
 metal column density, $N_{\rm cold} Z_{\rm cold}$, separated by
 $N_{\rm cold} Z_{\rm cold} = 10^{21} \ [Z_\odot \ \rm cm^{-2}]$.
 The dotted curve is the total \lya LF without any decomposition.
 \textit{Bottom}: The \lya LFs of starburst galaxies are shown by thick
 curves, which are decomposed into the pre-outflow, outflow, and
 post-outflow phases.  
 The \lya LF of the quiescent galaxies is shown by the thin dashed line,
 and the thin dotted line is the total \lya LF.
 }
 \label{fig-z5p7}
\end{figure}

  We first compare the simplest model (the simply proportional model) to
  the cumulative \lya LF data at $z=5.7$ in Figure~\ref{fig-z5p7}.
  The best-fit is $f_0=0.60$, and this model overproduces the bright-end
  of the LAE LF.
  \placefigure{fig-z5p7}
  We see from the top panel of Figure~\ref{fig-z5p7} that the dominant
  component at the bright-end of the \lya LF is the starburst galaxies,
  while the quiescently star forming galaxies dominate at the faint end.
  If we try to adjust the model to the bright end, we must assume a much
  lower value of $\fesclya \ll 1$ than the observationally inferred
  value by Gawiser et al. (2006).  This could be one of the possible
  reasons that Le Delliou et al. (2005, 2006) obtained $\fesclya \ll 1$
  with the same treatment for \fesclya, though their model is
  completely independent of our own and a clear comparison is not easy.
 
  The middle and bottom panels show that the starburst galaxies in the
  pre-outflow phase having high metal column densities
  ($N_\mathrm{cold}Z_\mathrm{cold}\ge 10^{21}~\mathrm{Z_\odot~cm^{-2}}$)
  dominate in the largest luminosity range.  This is reasonable since
  the most efficient \lya photon production is expected during young
  stage of the starburst, and the galaxies in such phase are expected to
  be gas and metal rich.  This result indicates that the effect of interstellar
  dust extinction in \fesclya models would reduce the number of LAEs at
  the bright end, in a direction to a better agreement with
  observations.
 
\begin{figure}
 \epsscale{0.5} 
 \plotone{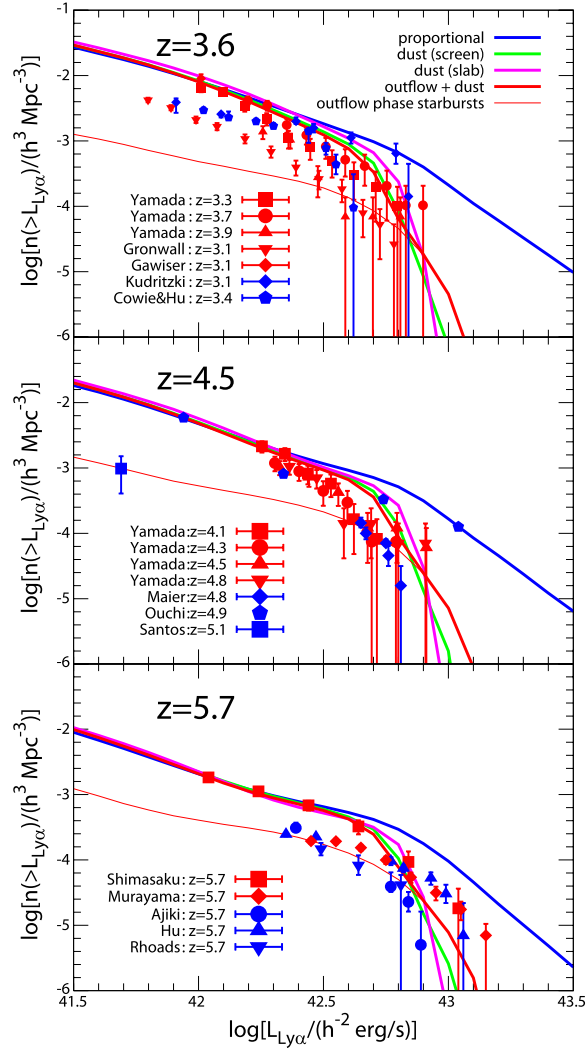} 
 \caption{ 
 The evolution of the cumulative \lya LFs with redshift at $z \lesssim 6$.  
 The curves show the model results, while the symbols with error bars
 are the observational data. 
 The four different models of $\fesclya$, i.e., the simply proportional, 
 dust (screen), dust (slab), and outflow$+$dust models are shown (see
 the top panel for the line markings).
 The thin solid curve is the contribution from the starbursts in the
 outflow phase in the outflow$+$dust model.
 The references for the data points that were not given in Figure
 \ref{fig-z5p7} are: Yamada et al. (2005), Gronwall et al. (2007),
 Gawiser et al. (2006), Kudritzki et al. (2000), Cowie \& Hu (1998),
 Maier et al. (2003), Ouchi et al. (2003), and Santos et al. (2004).
 }
 \label{fig-evolveLF}
\end{figure}

  We then show the dust models and the outflow$+$dust model 
  in Figure~\ref{fig-evolveLF}.
  \placefigure{fig-evolveLF}
  The fitted values of the model parameters are tabulated in
  Table~\ref{tab-freeparam}. 
  \placetable{tab-freeparam}
  As expected, the number of the luminous LAEs drastically decreases in
  the dust models.
  However, it can be seen that this effect is too strong, and the sharp
  drop of LAE number at the bright end does not match well to the
  observed \lya LFs.  
  In contrast, the outflow$+$dust model well reproduces the observed
  \lya LFs in the whole range of \lya luminosity shown in
  Figure~\ref{fig-evolveLF}.
  It should be noted again that the number of free model parameters is
  the same for the dust and outflow$+$dust models.
  LAEs at the bright-end are dominated by galaxies in the outflow phase,
  which is consistent with the observational indications of outflows.
  Furthermore, $\fesclya \sim 0.8$ has been assumed for the outflowing
  LAEs, which is consistent with a recent observational estimate
  for LAEs at $z\simeq 3.1$ by Gawiser et al. (2006).

 \subsection{Luminosity Functions at $z \gtrsim 6$ and IGM transmission}
 
\begin{figure}
 \epsscale{0.6}
 \plotone{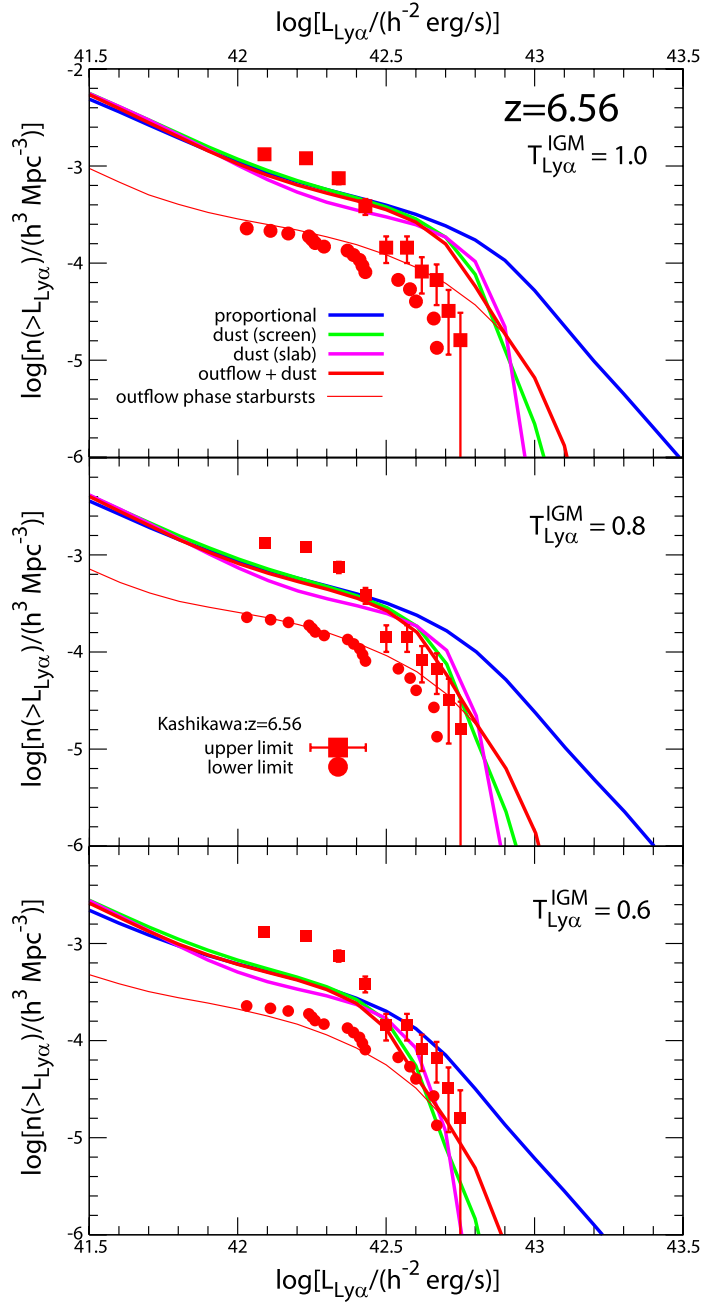}
 \caption{
 The cumulative \lya LFs at $z=6.56$. The line-markings are the
 same as Fig. \ref{fig-evolveLF}, but with different values of 
 the IGM transmission $\tlya =1.0,$ 0.8, and 0.6 for the top, middle,
 and bottom panels, respectively.
 The open squares are the observed data from the photometric sample
 of Kashikawa et al. (2006b) corrected for detection completeness (an
 upper limit), while the circles from the pure spectroscopic sample (a
 lower limit).
 }
 \label{fig-Taz6p5}
\end{figure}

\begin{figure}
 \epsscale{0.6}
 \plotone{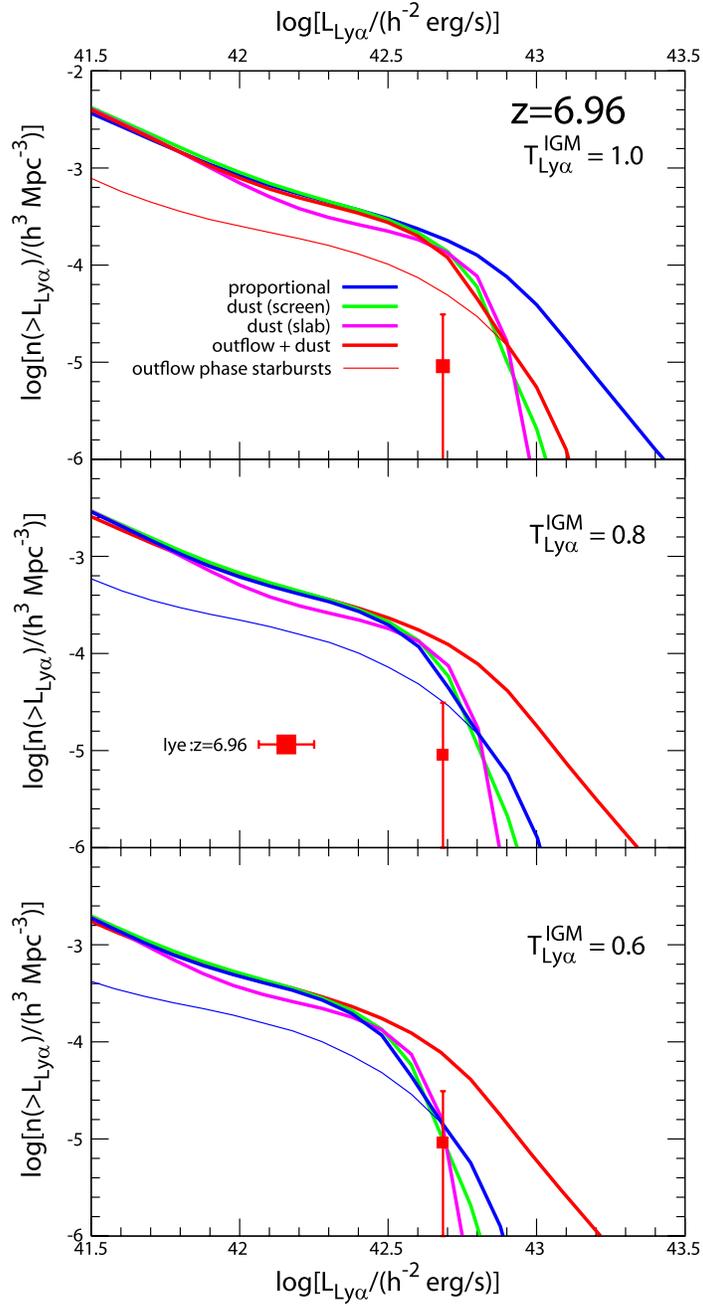}
 \caption{The same as Fig. \ref{fig-Taz6p5}, but at $z=6.96$. 
 The data is from Iye et al. (2006).
 }
 \label{fig-Taz6p9}
\end{figure}

  Turning now to the \lya LFs at redshift beyond 6 when the reionization
  is believed not to have ended yet.
  In contrast to the agreement at the lower redshifts, all \fesclya
  models assuming $\tlya = 1$ with the same parameters determined at
  $z=5.7$ overpredict the number of the LAEs at $z=6.56$ compared with
  the observational data reported by Kashikawa et al. (2006b), as
  presented in the top panel of Figure~\ref{fig-Taz6p5}.
  This discrepancy can be resolved if we adopt a simple prescription of
  luminosity-independent IGM transmission that is less than the unity
  ($\tlya \sim$ 0.6--0.8); the predicted \lya LFs with $\tlya =$
  0.8 and 0.6 are shown in the middle and bottom
  panels, respectively, in Figure~\ref{fig-Taz6p5}.
  \placefigure{fig-Taz6p5}
  We also show the predicted \lya LFs at an even higher redshift of $z=6.96$
  in Figure~\ref{fig-Taz6p9}.  
  \placefigure{fig-Taz6p9}
  Though the statistics is obviously insufficient, the most distant
  galaxy found so far (Iye et al. 2006) indicates $\tlya \lesssim$ 0.7. 
  We will discuss the implications for the reionization history from
  these results in \S~\ref{section:reionization}.

  \subsection{Interpretations for Intrinsic \lya LF Evolutionary
  Features}
 
\begin{figure}
 \epsscale{0.8}
 \plotone{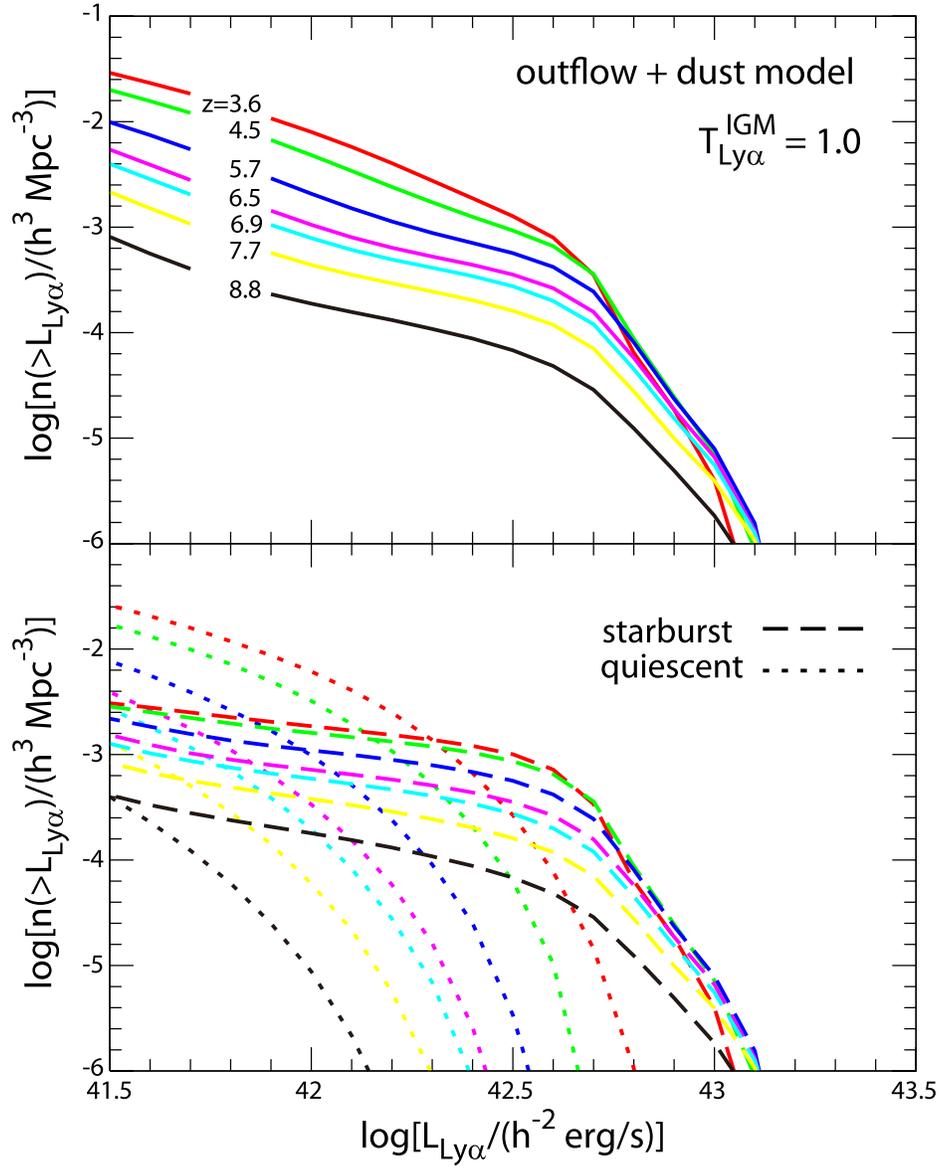}
 \caption{
 The evolution of the cumulative \lya LF in the outflow$+$dust model from
 redshift $z=3.6$ to $z=8.8$, assuming transparent IGM to \lya photons
 ($\tlya =1$).
 The contributions from starburst and quiescent galaxies are presented
 separately in the bottom panel.
 }
 \label{fig-intevolveLF}
\end{figure}

\begin{figure}
 \epsscale{1.0}
 \plotone{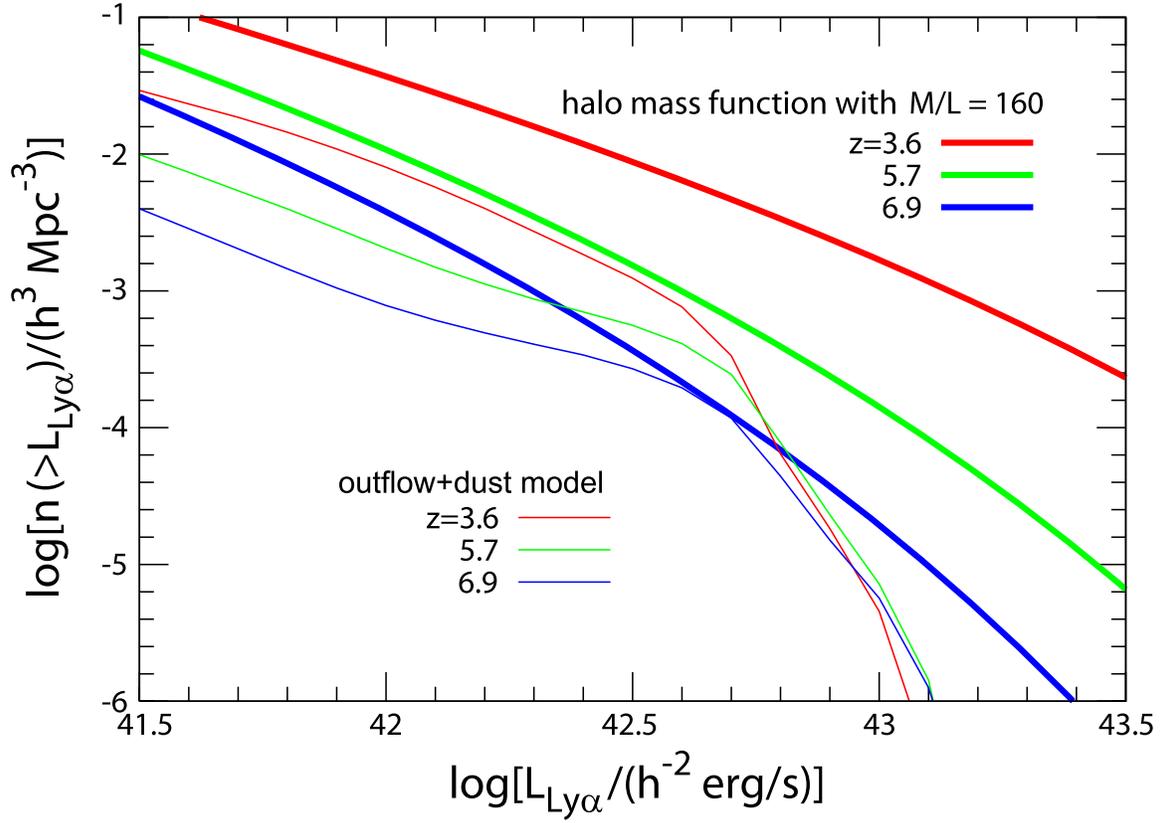}
 \caption{
 A comparison of the evolution of the cumulative \lya LF in the
 outflow$+$dust model (\textit{thin curves}) with that
 of the dark halo mass function (\textit{thick curves}) from redshift
 $z=3.6$ to $z=6.9$.
 The conversion factor of the halo mass into the \lya line luminosity,
 $M_\mathrm{DM}/L_\mathrm{Ly\alpha} = 160~M_\odot/L_\odot$, is
 chosen to fit the cumulative \lya LF at $z=6.9$.
 }
 \label{fig-MLcomp}
\end{figure}

  We show the intrinsic (i.e., $\tlya =1$) evolution of the cumulative
  \lya LF in the outflow$+$dust model from redshift $z=3.6$ to $8.8$ in
  Figure~\ref{fig-intevolveLF}.  It is also interesting to compare this
  evolution with that of the dark halo mass function, since a number of
  previous studies predicted LAE LF evolution based on the halo mass
  function.  This is shown in Fig. \ref{fig-MLcomp}, using the halo mass
  function given by Yahagi et al. (2004, hereafter YNY) that is used in
  the Mitaka model. Here, we adopt $M_\mathrm{DM} / L_\mathrm{Ly\alpha}
  = 160~{M_\odot / L_\odot}$, where $L_\odot$ is the bolometric solar
  luminosity.

  \placefigure{fig-intevolveLF}
 
  \placefigure{fig-MLcomp}
 
  A clear trend of the LAE LF evolution is that the characteristic
  luminosity where the LFs have a break, $L_{\rm br}$ (similar to $L_*$
  in the Schechter function), shows very little evolution in a wide range
  of redshift. This seems the primary reason that the LF is not very
  sensitive to the redshift at $L_{\rm Ly \alpha} \gtrsim L_{\rm br}$.
  On the other hand, the degree of the evolution at $L_{\rm Ly \alpha}
  \lesssim L_{\rm br}$ is similar to that of the dark halo mass
  function. Therefore, a mechanism that keeps the brightest
  Ly$\alpha$ luminosity almost constant against redshift would explain
  these trends.
 
  Such a mechanism is likely to be a combination of the dust extinction
  and outflow effect. The comparison of the dust and dust+outflow models
  with the simply proportional model (Figs. \ref{fig-z5p7} and
  \ref{fig-evolveLF}) indicates that galaxies more massive than a
  critical mass scale cannot become bright LAEs because of heavy
  extinction by a large amount of gas and metals, and/or because of the
  large gravitational potential that prevents an efficient outflow.
  Therefore, even if the number of galaxies more massive than the
  critical mass scale should increase significantly with time by
  hierarchical structure formation, the mass of the brightest LAEs does
  not significantly evolve. This interpretation can be tested by
  estimating the mass of LAEs in future observations.

 \section{Implications for Cosmic Reionization}
 \label{section:reionization}
 
The requirement of $\tlya < 1$ to reproduce the observed \lya LFs at $z
\gtrsim 6$ suggests that the IGM opacity for the \lya emission rapidly
increases beyond $z \sim 6$. From the value of \tlya estimated in our
analyses, we can estimate the IGM neutral fraction $\xhi \equiv n_{\rm
HI}/n_{\rm H}$.  Unfortunately, this procedure is highly complicated
because there are many physical processes that must be considered in
calculation of the attenuation factor of \lya line luminosity (Santos
2004; Dijkstra et al. 2007a).  Here, we apply the dynamic model with a
reasonable velocity shift of \lya line by 360 km/s redward of the
systemic velocity (the dashed curve of Figure 25 in Santos 2004).  
 
The reason for the choice of this particular 
model is that it predicts no
attenuation when $x_{\rm HI} = 0$.  Note that some other models of
Santos (2004) predict a significant attenuation even in the case of
$x_{\rm HI}=0$, due to the neutral gas associated with the host haloes
of LAEs. Choosing this model then means that we ascribe the
sudden strong evolution of the Ly$\alpha$ LF at $z \gtrsim 6$ only to
the absorption by pure IGM. We consider that this is a reasonable
assumption, since observations indicate that the escape fraction of
Ly$\alpha$ photons is about unity at least for LAEs at $z \sim 3$
(Gawiser et al. 2006). If LAEs at $z \sim 7$ are a similar population to
the low-$z$ LAEs, we do not expect significant absorption by neutral gas
physically associated to LAEs. It should also be noted that the
brightest LAEs in our model LFs are in the outflow phase, and we
may not expect a large amount of neutral gas around them.
 
For the range of $\tlya = 0.6-0.8$ at $z = 6.56$ estimated from the
outflow$+$dust model, we find $\xhi \sim 0.25-0.35$.  At $z = 6.9$, our
constraint of $\tlya \lesssim 0.7$ translates into $\xhi \gtrsim 0.3$.
On the other hand, it should also be kept in mind that if $z\sim 7$ LAEs
are surrounded by a significant amount of nearby neutral gas that is not
present for LAEs at $z \sim 3$, 
the estimate of $x_{\rm HI}$ as an average of
IGM in the universe could become lower than those derived here.

  Although the translation from \tlya into \xhi is model dependent, it
  should be noted that our model overpredicts the observed LAE LFs
  rather suddenly beyond $z \sim 6$, while it reproduces well the
  observed LAE LFs at $z \lesssim 6$ without invoking the
  absorption in IGM.
  We suggest that this is an evidence for significant absorption
  by neutral IGM at $z \gtrsim 6$, and it requires $\xhi$ of order
  unity.  This is broadly consistent with the recent constraints by
  other methods (Fan et al. 2006).  Here we emphasize that there has
  been no strong ``positive'' evidence for a considerable amount of IGM
  neutral hydrogen (\xhi of order unity) at $z \gtrsim 6$; quasar
  Gunn-Peterson tests give only a weak lower limit of $\xhi
  \gtrsim 10^{-3}$ (Fan et al. 2006) and the GRB 050904 gives only an
  upper limit of $\xhi \lesssim 0.6$ (Totani et al. 2006).
 
  For the future studies of reionization through LAE LFs, we plot the LF
  evolution of the outflow$+$dust model from $z=3.6$ to $z=8.8$ in
  Figure~\ref{fig-intevolveLF} with $\tlya = 1$.  The redshifts of $7.7$
  and $8.8$ correspond to the next windows that are relatively free of
  bright OH lines at wavelengths of 1.06 \micron~ and 1.19 \micron,
  respectively.  There are already several pioneering works, such as
  Willis \& Courbin (2005), Cuby et al. (2007) and Stark et al. (2007b).
  Moreover, other wide-field imagers with narrow-band filters to search
  the redshifted \lya emission at these wavelengths will be available on
  4 m or 8-10 m class telescopes in the near future (see, e.g.  Cuby et
  al. 2007, for more detail).  Our predicted LAE LF evolution would be
  useful for the planning of the future LAE surveys, and it can also be
  used as a guide to derive $\tlya$ from the observed LAE
  LFs\footnote{The numerical data of the LF evolution is available on
  request to the authors.}.

 \section{Conclusions}
 \label{section:conclusions} 
 
 We constructed a new theoretical model for Lyman alpha emitters at
 high-$z$ based on a hierarchical clustering model of galaxy formation
 (the Mitaka model), taking into account physical effects of dust
 absorption and galaxy-scale outflow in calculation of \lya photon
 escape fraction ($\fesclya$) from host galaxies.  We introduced just
 two new parameters based on some physical considerations to include
 these two effects, while we kept unchanged the original model
 parameters of the Mitaka model that have been tuned to reproduce a
 number of observations of local galaxies.
 
The observed \lya LFs of LAEs at several redshifts in $z \lesssim 6$ are
 reproduced well by our model assuming completely ionized IGM (the IGM
 transmission $\tlya =1$). Especially, it has been known that the LAE
 LFs show very little evolution in $z \sim 3$--6 compared with the dark
 halo mass function, and our model can naturally explain this by the
 above effects newly included in this work.  The escape fraction of
 Ly$\alpha$ photons from LAEs in our model is $\fesclya \sim 0.8$, which
 is consistent with a recent observational estimate. Moreover, our model
 predicts that the bright-end of the \lya LF is dominated by the
 starburst galaxies in the phase of galaxy-scale outflow driven by the
 supernova feedback. This is also consistent with the observational
 results for local and high-$z$ galaxies with strong \lya emission.  On
 the other hand, the simple model with a constant and universal \fesclya
 requires $\fesclya \ll 1$ in order not to overproduce the brightest
 LAEs. This is consistent with a previous theoretical study that
 invoked the same assumption, but the required
 $\fesclya$ value is considerably lower than the observed values.

 In contrast to the success at $z \lesssim 6$, our model overpredicts
 the LAE LF beyond $z \gtrsim 6$, and we interpret this as a result of
 \lya attenuation by neutral hydrogen in IGM.  We found that a simple
 prescription of luminosity-independent $\tlya$ is sufficient to explain
 the observed LAE LFs; the LF at $z=6.56$ is reproduced with $\tlya
 =0.6-0.8$, and that at $z=6.9$ indicates $\tlya \lesssim 0.7$.  Though
 it is not straightforward to derive the IGM neutral fraction \xhi from
 these results, the discrepancy in the case of $\tlya = 1$
 appears rather suddenly beyond $z \gtrsim 6$ compared with the results
 in $z \lesssim 6$.  Therefore we suggest that this gives an evidence
 for a rapid evolution of \xhi from $\xhi \ll 1$ at $z \lesssim 6$ to a
 value of order unity at $z \gtrsim 6$.  This positive evidence for a
 significant abundance of neutral hydrogen in IGM is complementary to
 the constraints obtained by other methods based on quasar and GRB spectra.
 
 Our theoretical model for the LAE LF evolution would be
 helpful for planning a LAE survey and for interpretation of the
 observed \lya LF at even higher redshifts in the future studies.
 Further investigation of various properties (e.g., spectral energy
 distribution, mass, or spatial clustering) of LAEs in our model and
 comparison with observations will be done in our future studies.
 
 
\acknowledgments
 
We would like to thank the referee for useful comments.  We would also
like to thank Masanori Iye, Nobunari Kashikawa, Kazuaki Ota, Kazuhiro
Shimasaku, and Caryl Gronwall for providing their observational data,
and Masataka Ando, Akio Inoue, Yuichi Matsuda, Tohru Nagao, Koji Ohta,
and Masami Ouchi for useful discussions. 
This work was supported in part by the Grant-in-Aid for Scientific
Research (16740109 for T.T.) and for the 21st Century COE "Center for
Diversity and Universality in Physics" from the Ministry of Education,
Culture, Sports, Science and Technology (MEXT) of Japan.  
M.A.R.K. has been supported by the JSPS (Japan Society for the Promotion
of Science) Research Fellowships for Young Scientists.
 
 

\clearpage
 
\begin{deluxetable}{cccc}
 \tabletypesize{\scriptsize}
 \tablewidth{0pt}
 \tablecolumns{3}
 \tablecaption{
 The best-fit model parameters of \lya photon escape fraction.
 \label{tab-freeparam}
 }
 
 \tablehead{
 \colhead{\fesclya model} &
 \colhead{\phantom{sssssssssss}$f_0$\phantom{ssssssssss}} &
 \colhead{$\left(N_\mathrm{cold}Z_\mathrm{cold}
 \right)_0~\left[10^{20}~Z_\odot
 ~\mathrm{cm^{-2}}\right]$}
 }
 
 \startdata
 simply proportional....... & $0.60^{+0.05}_{-0.04}$ & \nodata & \\
 dust (screen).................. & $0.67^{+0.04}_{-0.05}$ & $6.0^{+2.7}_{-2.0}$ \\
 dust (slab)..................... & $0.80^{+0.05}_{-0.05}$ & $2.4^{+0.8}_{-0.7}$\\
 outflow$+$dust................ & $0.68^{+0.05}_{-0.08}$ & $4.2^{+1.5}_{-1.5}$
 \enddata
 
 \tablecomments{The errors are statistical 1$\sigma$ as a result of the
 fit to the \lya LF of the LAEs at $z=5.7$ reported by Shimasaku et
 al. (2006).
 The screen dust extinction is adopted in the pre-outflow phase of the
 outflow$+$dust model.}
 
\end{deluxetable}
 
\clearpage
 
\end{document}